\documentclass[preprint,preprintnumbers,amsmath,amssymb]{revtex4}
\usepackage{graphicx}
\usepackage{dcolumn}
\usepackage{bm}
\begin{document}
\title {Topology, mass and Casimir energy}
\author{ N.~Ahmadi$^{a}$\footnote{
Electronic address:~nahmadi@ut.ac.ir} and M.~Nouri-Zonoz $^{a,b}$ \footnote{
Electronic address:~nouri@theory.ipm.ac.ir}}
\address{$^{a}$ Department of Physics, University of Tehran, North Karegar Ave., Tehran 14395-547, Iran \\
$^{b}$ Institute for studies in theoretical physics and mathematics, P O Box 19395-5531 Tehran, Iran.}
\begin{abstract}
The vacuum expectation value of the stress energy tensor for a massive scalar field with arbitrary coupling
in flat spaces with non-trivial topology is discussed. We calculate the Casimir energy in these spaces 
employing the recently proposed {\it optical approach} based on closed classical paths. The evaluation of 
the Casimir energy consists in an expansion in terms of the lengths of these paths. We will show how 
different paths with corresponding weight factors contribute in the calculation. The optical approach 
is also used to find the mass and temperature dependence of the Casimir energy in a cavity and it is 
shown that the massive fields cannot be neglected in high and low temperature regimes. The same approach 
is applied to twisted as well as spinor fields and the results are compared with those in the literature. 
\end{abstract}
\maketitle
%%%%%%%%%%%%%%%%%%%%%%%%%%%%%%%%%%%%%%%%%%%%%%%%%%%%%%
\section{Introduction}
Investigations of the different aspects of quantum field theory in spaces with non-trivial topology \cite{1} 
may be interesting for several reasons. First, it is a possibility that the early Universe was in fact multiply 
connected, so the combined effect of topology and curvature is likely to have been of considerable importance 
in the early stages of the evolution of the universe. Second, in theories with extra dimensions such as the 
well known Kaluza-Klein theory, extra dimensions may need to be compactified providing the space with non-trivial 
topology. Furthermore field theory in spaces with non-trivial topology (the one-torus compactified space) is 
quite similar to the field theory at non-zero temperature. The Casimir effect on spaces with non-trivial topology 
has been investigated in different contexts (\cite{2} and references therein). Most of the previous studies are 
concerned with one compact dimension (representing imaginary time or a compact spatial dimension) in either a 
curved or flat spacetime \cite{3}. The issue of obtaining the right order of magnitude for the possible contribution 
of the Casimir energy density to the cosmological constant, corresponding to a free scalar field, is considered in  
toroidal universes with $p$ large and $q$ small dimensions \cite{4}. Since the Vacuum polarization appears in 
manifolds with non-trivial topologies as well as in bounded domains with boundary conditions, the question of 
the definition of the vacuum energy is a most fundamental one in any quantum field theory. On the other hand 
imposing (mathematical) boundary conditions on physical quantum fields turns out to be a highly non-trivial 
act. The difference between the zero-point energies corresponding to the scalar field when boundary condition 
is enforced  and that in the absence of any boundary condition, diverges as the boundary is approached. Integrating 
over all space yields an infinite vacuum energy per unit area of the boundary surface. Since the infinite terms 
do not depend on the distance between the plates, these divergences do not usually show up in the calculation 
of Casimir force. To obtain the finite results through the use of a suitable regularization, many efforts 
have been done to identify and separate the singularities by physically meaningful 
cutoffs \cite{5}. The whole argument is based on boundaries that constrain {\it all} modes of the fluctuating field, 
introducing divergences into physical observables. Therefore one needs a realistic boundary condition through 
a high energy cutoff that limits the fluctuations and thereby introduces a high energy scale into the theory. 
Modes with energy below this scale obey a boundary condition while modes with energies at or above this scale 
do not. Hence the boundary condition applies to the physics at energies much lower than this scale where there 
is no need to reference the scale, as it was the case in the Casimir's original problem. In this paper we examine 
the Casimir-type effects in four dimensional spacetimes with $R(time)\times\Sigma(space)$ topology, using an 
approach introduced recently \cite{6} for an approximate calculation
of Casimir energy in generic geometries like the Casimir pendulum and a sphere opposite a plane. Although an 
approximation, it leads to exact results for flat geometries. It is specially useful when the normal modes 
cannot be determined. The basic idea is to employ the regularization method of {\it point splitting} through 
an expansion of the Casimir energy in terms of the classified, closed geodesic paths and then  identify those 
classes that contribute to the Casimir effect. The expansion is given in terms of the lengths of the paths and 
the paths with zero-length (corresponding to the high frequency components of the field) introduce divergences 
as they only probe the geometry in the immediate vicinity of the spacetime point of interest and in this 
restricted neighbourhood the topology does not change.\\  
Including a massive field in the Casimir energy calculations is the simplest generalization of the usual, massless 
scalar field calculations. Casimir-type effects in the presence of massive fields have been studied in \cite{7} 
and \cite{8} for both scalar and spinor fields. There it has been found that a non-zero field mass always decreases 
the magnitude of the energy density. Here we will encounter cases in which massive fields act in the opposite direction. 
The outline of the paper is as follows, in Sec. II the stress-energy tensor is written in terms of the trace of the 
Hadamard function which is the basic object from which $\left\langle 0\left|T_{\mu\nu}\right|0\right\rangle$ is 
constructed by differentiation. We will review the Casimir energy calculations between parallel plates as an 
example of the use of the optical approach in section III. Some applications of this approach in flat four dimensional 
manifolds, either orientable or non-orientable, are studied in section IV. Divergences due to the massive fields at 
low and high temperature limit are also discussed in this section. Studying spinor fields in section V, we show that 
the optical method can yield exact results. Conclusions and summary are discussed in the last section.
%%%%%%%%%%%%%%%%%%%%%%%%%%%%%%%%%%%
\section{Stress-energy tensor} 
The action functional for a scalar field $\Phi$ in a curved spacetime is
\begin{equation}
 S\left[\Phi\right]=\frac{1}{2}\int\sqrt{-g}\left( \Phi_{;\mu}\Phi^{;\mu}-\varsigma {\cal R}\Phi^{2}-m^{2}\Phi^{2}\right)d^{4}x\label{1}
\end{equation}
where $g$ is the determinant of the background metric $g_{\mu\nu}$, $\cal{ R}$ is the curvature scalar and $\varsigma$ is the gravitational coupling. Varying $\Phi$ infinitesimally, we obtain the scalar field equations $\left[\square_{x}+M^2+\varsigma \cal {R}\right]\Phi=0$. The classical scalar stress tensor density is given by \cite{9}
\begin{eqnarray}
 T_{\mu\nu} &=&\left(1-2\varsigma\right)\Phi_{;\mu}\Phi_{;\nu}+\left(2\varsigma-\frac{1}{2}\right)g_{\mu\nu}\Phi_{;\alpha}\Phi^{\alpha}-2\varsigma\Phi_{;\mu\nu}\Phi+\frac{1}{2}\varsigma g_{\mu\nu}\Phi\square\Phi\nonumber\\
 &&-\varsigma\left({\cal {R}}_{\mu\nu}-
\frac{1}{2}g_{\mu\nu}{\cal {R}}+\frac{3}{2}\varsigma {\cal {R}} g_{\mu\nu}\right)\Phi^{2}+\frac{1}{2}\left(1-3\varsigma\right)m^{2}g_{\mu\nu}\Phi^{2}
\label{2}\end{eqnarray}                        
The transition from classical to quantum fields is made by replacing the classical field $\Phi$ by a field operator. We then note that eq. (\ref{2}) is constructed from the products of the field operators or their derivatives at the same spacetime point. These quantities diverge when their vacuum expectation value is taken. To avoid these divergent quantities, one operator in each product is moved to a nearby point. The point separated vacuum expectation value of the energy-momentum tensor is then expressed in terms of the so-called Hadamard function $G^{(1)}\left(x,\widetilde{x}\right)$ as,
 \begin{eqnarray}
 &&\left\langle 0\left| T_{\mu\nu}\right|0\right\rangle=\frac{1}{2}\lim_{\widetilde{x}
\rightarrow x}\left[\left(1-2\varsigma\right)\nabla_{\mu}\widetilde{\nabla}_{\nu}+\left(2\varsigma-\frac{1}{2}\right)
g_{\mu\nu}\nabla_{\alpha}\widetilde{\nabla}^{\alpha}-2\varsigma\nabla_{\mu}\nabla_{\nu}\right.\nonumber\\
&&\left.+\frac{1}{2}\varsigma g_{\mu\nu}\nabla_{\alpha}\nabla^{\alpha}-\varsigma\left({\cal {R}}_{\mu\nu}-\frac{1}{2}{\cal{R}} g_{\mu\nu}+\frac{3}{2}\varsigma {\cal {R}}g_{\mu\nu}\right)  -\frac{1}{2}\left(1-3\varsigma\right)m^{2}g_{\mu\nu}\right]G^{(1)}\left(x,\widetilde{x}\right)\label{3}
 \end{eqnarray}
where $G^{(1)}\left(x,\widetilde{x}\right)$ is defined by $G^{(1)}\left(x,\widetilde{x}\right)=\left\langle 0\left|\left\{\Phi\left(x\right),\Phi\left(\widetilde{x}\right)\right\}\right|0\right\rangle$, and satisfies  
  \begin{equation}
  \left[\square_{x}+m^2+\varsigma {\cal {R}}\right] G^{(1)}\left(x,\widetilde{x}\right)=0\label{4}.
  \end{equation}
Because of the difficulties in calculations of quantum field theory in a general curved spacetime, it is very natural to deal with some specific spaces. To study the effect of topology on Casimir energy, we relax the curvature effect by restricting ourselves to flat spacetimes with $R\times\Sigma$ topology whose metric is given by
\begin{equation}
    \left( ds\right) ^2=\left( dx^0\right) ^2-\left( dx^1\right) ^2-\left(
dx^2\right) ^2-\left( dx^3\right) ^2.           
\label{5}\end{equation}                                                            
 For the Minkowski space $G^{(1)}\left(x,\tilde{x}\right)$ is given by \cite{10}
\begin{equation}
    G^{(1)}\left(x,\widetilde{x}\right)=\frac m{2\pi ^2\sqrt{2\sigma }}\Theta (2\sigma
)K_1(m\sqrt{2\sigma })+\frac m{4\pi \sqrt{-2\sigma }}\Theta (-2\sigma )I_1(m%
\sqrt{-2\sigma })            
\label{6}.\end{equation}            
In which $\Theta$ is the step function and $K_1$ and $I_1$ are the modified Bessel functions of the first and second kind, respectively. The quantity 
$\sigma\left(x,\tilde{x}\right)=\frac {1}{2}g_{\alpha \beta }\left(x^\alpha -\tilde{x}^\alpha \right)\left(x^\beta -\widetilde{x}^\beta \right)$ 
is a biscalar equal to one half the square of the geodesic distance between $x$ and $\widetilde{x}$ \cite{11}. As discussed in the introduction, divergences occurring in the stress-energy tensor calculations are due to the high frequency components of the field. In Minkowski space quantum field theory, these divergences are simply discarded using the normal ordering prescription for field operators. On the other hand one may employ an ultraviolet regulator function $e^{-k/\Lambda}$, to cut off the ultraviolet divergences, and then take the difference between the values of $\left\langle 0\left| T_{\mu\nu}\right|0\right\rangle$ in the presence of the boundary or the topological condition and in the absence of them, letting $\Lambda\rightarrow\infty$ at the end of the calculation. Alternatively, in the Green function approach, one first subtracts the unbounded Minkowski space expression for $G^{(1)}\left(x,\tilde{x}\right)$ from the same function evaluated in the topology of interest, and then takes the limit $\widetilde{x}\rightarrow x$. In this paper using the latter approach we expand  Green functions in terms of all geodesic paths from $x$ to $\tilde{x}$, classified and labeled by the family index $n$ through their endpoints $x_n$.  The lengths of these paths contribute with equal or different weight factors, $M_{n}$, 
\begin{equation}
G^{(1)}\left(x,\tilde{x}\right)=\sum_{n}M_{n}G^{(1)}\left(x,\tilde{x}_{n}\right).
\end{equation} 
In this approach, it will become clear that the divergences arise due to arbitrarily short length paths which could be isolated and discarded from the outset, leading to the infinity-free Casimir energy calculations. 
%%%%%%%%%%%%%%%%%%%%%%%%%%%%%% 
\section{Casimir energy between the parallel plates}
\subsection{Ideal parallel conductors}
In this section, using the optical approach, we review the Casimir energy corresponding to a massive scalar field in the most simple configuration i.e. a rectangular parallelepiped with ideal walls of dimension $a,b$ and $c$ with $a \ll b,c$. We employ Dirichlet condition at the boundaries i.e., 
\begin {equation}
\Phi\left(x^{0},0,x^{2},x^{3}\right)=\Phi\left(x^{0},a,x^{2},x^{3}\right)=0
\label{8}\end{equation}
 The stress-energy tensor is given by
  \begin{eqnarray}
  T_{\mu\nu} &=&\left(1-2\varsigma\right)\Phi_{;\mu}\Phi_{;\nu}+\left(2\varsigma-\frac{1}{2}\right)g_{\mu\nu}\Phi_{;\alpha}\Phi^{;\alpha}\nonumber\\
  &&-2\varsigma \Phi_{;\mu\nu}\Phi+\frac{1}{2}\varsigma g_{\mu\nu}\Phi\Box\Phi+\frac{1}{2}\left(1-3\varsigma\right)m^{2}g_{\mu\nu}\Phi^{2}.
  \label{9}\end{eqnarray}                 
Using the field equations along with eq. (\ref{8}), vacuum expectation value of $T_{\mu\nu}$ in a flat spacetime could be expanded in terms of geodesic paths from $ x$ to $\widetilde{x}_{n}$,
\begin{eqnarray}
\left\langle  0\left|T_{\mu\nu}\right|0\right\rangle&=&\frac{1}{2}\lim_{\tilde{x}\rightarrow x}
\left[\left(1-2\varsigma\right)\nabla_{\mu}\widetilde{\nabla}_{\nu}+\left(2\varsigma-\frac{1}{2}\right)g_{\mu\nu}\nabla_{\alpha}\tilde{\nabla}^{\alpha}\right.\nonumber\\&&\left.-2\varsigma\nabla_{\mu}\nabla_{\nu}+2\varsigma g_{\mu\nu}\nabla_{\alpha}\nabla^{\alpha}+\frac{1}{2}m^{2}g_{\mu\nu}\right]\sum_{n}G^{(1)}\left(x,\widetilde{x}_{n}\right). 
\label{10}\end{eqnarray}                                                                                    Every geodesic path contributes to $G^{(1)}\left(x,\tilde{x}\right)$ and it is clear that $G^{(1)}\left(\sigma\right)$ and its derivatives diverge as $\sigma_{n}$ tends to zero. These paths fall into classes $C_{n}$, which have $n$ points on the boundaries. The length of the geodesic path that starts from $x$ and arrives at $\widetilde{x}$ after $n$ reflections from the boundary is shown by $l_{n}\left(x,\tilde{x}\right)$. These paths are closed in the limit $\widetilde{x}\rightarrow x$.
By imposing the above boundary conditions on the surface located at $x^{1}= 0$ normal to the closed direction, the geodesic paths will be reflected on arriving at the surface. Dirichlet boundary condition on the plates is implemented by the reflection factor (-1). The generic closed path starting at $x$, reaches $x_{n}$ after $n$-reflections from the boundary, where in the Cartesian coordinates $x_{n}$ is given by
\begin{equation}
\left( x^{0}_{n},x^{1}_{n},x^{2}_{n},x^{3}_{n}\right)=\left\{
\begin{array}{c}
\left( x^0,x^1+n_{e}a,x^2,x^{3}\right)\\
\left( x^0,3x^1+\left(n_{o}-1\right)a,x^2,x^3\right)\\
\left( x^0,-x^1+\left(n_{o}+1\right)a,x^2,x^3\right)
\end{array}
\right.
\label{11}\end{equation}
%%%%%%%%%%%%%%%%%%%%%%%%%%
\begin{figure}
   \centering
    \includegraphics{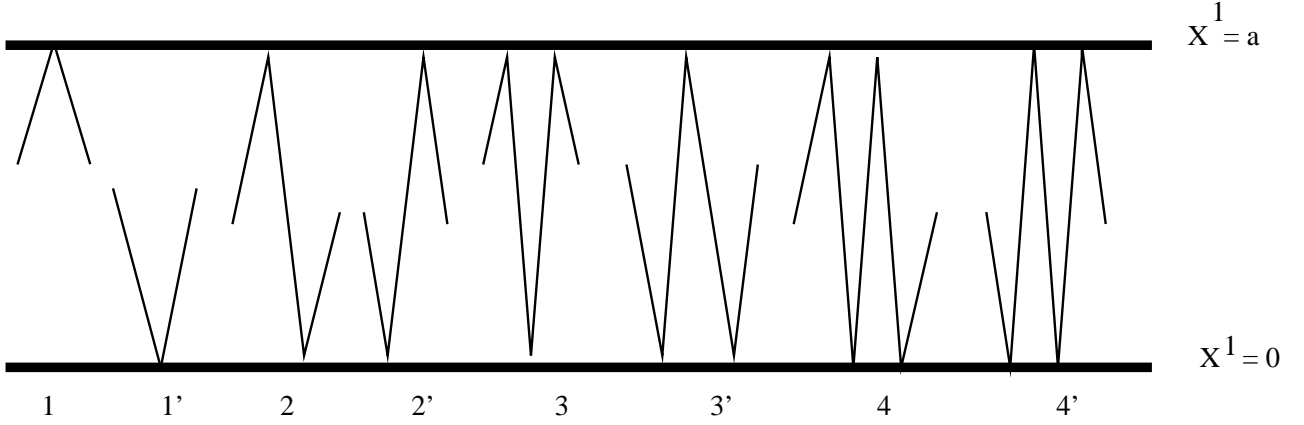}
 \label{fig:fig1.eps}\caption{Different geodesic closed paths. The initial and final points, which coincide, have been separated so that the paths could be identified.}
\end{figure}
%%%%%%%%%%%%%%%%%%%%%%%%%%%
in which $n_{e}$ and $n_{o}$ are even and odd integers respectively. It is convenient to separate the geodesic paths into \emph{odd} and \emph{even} paths, according to their number of reflections from the plates. Some of these paths are shown in Fig. 1. As it is seen, the two paths $2$ and  $2^{\prime}$ with two reflections are distinct, so the multiplicity factor, $M_{n}$ , is two for even paths, and one for odd ones. It is shown that odd and even paths differ in their contribution to the Casimir effect \cite{12}. The contribution of the even paths in the vacuum energy density $\left\langle0\left|T_{00}\right|0\right\rangle$ being finite, results in the Casimir force, while the odd paths lead to a divergent contribution. The half squared geodesic distance for a path with $n_{e}=2r$ reflections is
\begin{equation}   \sigma_{2r}=\frac{1}{2}\left[\left(x^{0}-\tilde{x}^{0}\right)^2-\left(x^{1}-\widetilde{x}^{1}-2ra\right)^2-\left(x^{2}-\tilde{x}^{2}\right)^2-\left(x^{3}-\tilde{x}^{3}\right)^2\right]
\label{12},\end{equation}                                
and its contribution to $\left\langle0\left|T_{00}\right|0\right\rangle$ is 
\begin{eqnarray}
\left\langle 0\left|T_{\mu\nu}\right|0\right\rangle_{n_{e}=2r}&=&\frac{1}{2}\lim_{\tilde{x}
\rightarrow{x}}M_{2r}\left[\left(1-2\varsigma\right)\nabla_{\mu}\tilde{\nabla}_{\nu}+\left(2\varsigma-\frac{1}{2}\right)g_{\mu\nu}\nabla_{\alpha}\tilde{\nabla}^{\alpha}\nonumber \right. \\ 
&&\left. -2\varsigma\nabla_{\mu}\nabla_{\nu}+2\varsigma  g_{\mu\nu}\nabla_{\alpha}\nabla^{\alpha}+\frac{1}{2}m^{2}g_{\mu\nu}\right] G^{(1)}\left(\sigma_{2r}\right).
\label{13}\end{eqnarray}                                       
Since $\tilde{\nabla}_{\mu}\sigma_{n_{e}}=-\nabla_{\mu}\sigma_{n_{e}}$\footnote{Note that the quantity $\sigma\left(x,\tilde{x}\right)$ is a biscalar, i.e. a function which transforms like the product of two scalars, one at each space-time point and the tangent vector at $x$ must be parallel transported to $\tilde{x}$ along the geodesic connecting  $x$ and $ \tilde{x}$ and then compared with that at $\tilde{x}$.}, the contribution from even paths is given by
  \begin{eqnarray}
\left\langle0\left|T_{00}\right|0\right\rangle_{even}&=&
   -\lim_{\tilde{x}\rightarrow x}\sum_{r=1}^{\infty }\nabla_{0}\nabla_{0}G^{(1)}\left(\sigma_{2r}\right)\nonumber\\
   &=&\frac{m^4}{2\pi^2}\lim_{\tilde{x}\rightarrow x}\sum\limits_{r=1}^\infty\left[\frac{K_2\left(A_{2r}\right)}{\left(A_{2r}\right)^2}-m^2\frac{K_3\left(A_{2r}\right)}{\left(A_{2r}\right)^3}\left(\nabla_{0}\sigma_{2r}\right)^2\right],
\label{14}\end{eqnarray}
\noindent with $A_{2r}=m l_{2r}=2mar$. Note that the paths in class $C_0$ $(r=0)$ are excluded in the above summation as they lead to divergences. Since the interval between two endpoints is spacelike, that is $\sigma_{n}>0$, only the first term in (6) concerned us in arriving at (\ref{14}). In coincidence limit, $\tilde{x}\rightarrow x$, we have
\begin{equation}
\left\langle0\left|T_{00}\right|0\right\rangle_{even}=\frac{m^2}{8\pi^2} \sum\limits_{r=1}^{\infty}\frac{K_2\left(2mra\right)}{r^{2}a^{2}}.
\label{15}\end{equation}                                                                               
Energy per unit area is obtained by integrating over the distance between the plates and is given by $\epsilon_{even}=\frac{m^2}{8\pi^{2}a} \sum\limits_{r=1}^{\infty}\frac{K_2\left(2mra\right)}{r^{2}}$. This is a rapidly convergent series where the main contribution comes from the first term. Note that this result is valid for an arbitrary value of $\varsigma$. In the limit $ma \ll 1$, we obtain 
\begin{equation}
\epsilon=\frac{1}{16\pi^{2}a^{3}}\sum\limits_{r=1}^{\infty}\frac{1}{r^4}= \frac{1}{16\pi^{2}a^{3}}\zeta\left(4\right)=\frac{\pi^2}{1440a^3}.
\label{16}\end{equation}                                                                    
In the limit $m\rightarrow 0$, the first term includes $90/\pi^4\approx0.92$ of the whole series.
Next consider the odd paths. There are two families in each $C_{2r+1}$ class illustrated in Fig. 1. One family begins its path with the first reflection from the lower plate while the primed family starts with reflections from the upper one. Half squared geodesic length for a path in class $C_{2r+1}$ is given by 
   \begin{equation}\left\{ \begin{array}{l}   \sigma_{2r+1}\\\sigma^{\prime}_{2r+1}\end{array}\right.=\frac{1}{2}\left[\left(x^{0}-\tilde{x}^{0}\right)^2-\left(\begin{array}{c}\left(x^{1}-3\tilde{x}^{1}-2ra\right)^2\\\left(x^{1}+\tilde{x}^{1}-2\left(r+1\right)a\right)^2\end{array}\right)-\left(x^{2}-\tilde{x}^{2}\right)^2-\left(x^{3}-\tilde{x}^{3}\right)^2\right].      \label{17}\end{equation}        
For the tangent vectors we have
\begin{equation}     
\tilde{\nabla}_{\mu}\sigma_{2r+1}=-\nabla_{\mu}\sigma_{2r+1}-2\delta_{\mu,1}\nabla_{1}\sigma_{2r+1}\qquad
\tilde{\nabla}_{\mu}\sigma^{\prime}_{2r+1}=-\nabla_{\mu}\sigma^{\prime}_{2r+1}+2\delta_{\mu,1}\nabla_{1}\sigma^{\prime}_{2r+1}
\label{18}\end{equation}
In the coincidence limit, the lengths of the paths in class $C_{2r+1}$ range from $2ra$ to $2\left(r+1\right)a$ as $x^1$ varies from zero to $a$. The paths with $n_{o}=2r+1$ reflections contribute in energy density as
\begin{eqnarray}                                                                                            &&\left\langle0\left|T_{00}\right|0\right\rangle _{2r+1}                                           \nonumber\\ 
&=&\frac{1}{2} \lim_{\tilde{x} \rightarrow{x}}\left\{\left[-\nabla_{0}\nabla_{0}-\left(4\varsigma-1\right)\nabla_{1}\nabla_{1}\right]G^{(1)}\left(\sigma_{2r+1}\right)+ \left[-\nabla_{0}\nabla_{0}+\left(4\varsigma-1\right)\nabla_{1}\nabla_{1}\right]G^{(1)} 
\left(\sigma^{\prime}_{2r+1}\right)\right\}                                                          \nonumber \\
&=&\frac{m^4}{4\pi^2}\lim_{\tilde{x}\rightarrow x}\left\{2\left(1-2\varsigma\right)\frac{K_{2}\left(A_{2r+1}\right)}{A_{2r+1}^2}+4\varsigma\frac{K_{2}\left(A_{2r+1}^{\prime}\right)}{{A_{2r+1}^{\prime}}^2}\right.\nonumber\\ 
&&-m^2\frac{K_{3}\left(A_{2r+1}\right)}{A_{2r+1}^3}\left[\left(\nabla_{0}\sigma_{2r+1}\right)^2-\left(4\varsigma-1\right)\left(\nabla_{1}\sigma_{2r+1}\right)^2\right]                            \nonumber\\
&&\left.-m^2\frac{K_{3}\left(A_{2r+1}^{\prime}\right)}{{A_{2r+1}^{\prime}}^3}\left[\left(\nabla_{0}\sigma_{2r+1}^{\prime}\right)^2-\left(4\varsigma-1\right)\left(\nabla_{1}\sigma_{2r+1}^{\prime}\right)^2\right]\right\}.
\label{19}
\end{eqnarray}
where $A_{2r+1}=ml_{2r+1}$ and  $A^{\prime}_{2r+1}=ml^{\prime}_{2r+1}$. In coincidence limit, we have
\begin{eqnarray}
\left\langle 0\left|T_{00}\right|0\right\rangle_{2r+1}&=&\frac{m^4}{2\pi^2}\left\{\left(1-2\varsigma\right)\frac{K_{2}\left(2m\left|{x^1+ra}\right|\right)}{2m^{2}\left(x^1+ra\right)^2}+2\varsigma\frac{K_{2}\left(2m\left|x^1-\left(r+1\right)a\right|\right)}{2m^2\left(x^1-\left(r+1\right)a\right)^2}\right.                                                   \nonumber\\ &&\left.+\left(1-4\varsigma\right)\left[\frac{K_{3}\left(2m\left|x^1+ra\right|\right)}{2m\left(x^1+ra\right)}-\frac{K_{3}\left(2m\left|x^1-\left(r+1\right)a\right|\right)}{2m\left(x^1-\left(r+1\right)a \right)}\right]\right\}.
\label{20}
\end{eqnarray}                            
In the above expression we have set $z=  \left| x^{1}+ra\right|$ or  $\left|x^{1}-\left(r+1\right)a\right|$. So the contribution of eq. (\ref{20}) to Casimir energy per unit area is
\begin {equation}
\epsilon_{2r+1}=\int\limits_{ra}^{(r+1)a}\left\langle 0\left|T_{00}\right|0\right\rangle dz.
\label{21}
\end{equation}
Using the behavior of the function $K_{\nu}\left(z\right)$ for small values of its argument and the positive real part of $\nu$ \cite{14} 
 \begin{equation}
 \lim_{z\rightarrow 0}K_{\nu}\left(z\right)\approx\frac{1}{2}\Gamma\left(\nu\right)\left(\frac{z}{2}\right)^{-\nu},
 \label{22}\end{equation}                                                                                   it is easily seen that the only divergent contribution comes from the paths in class $C_{1}$ ($r=0$) at  the lower limit of (\ref{21}). It does not depend on relative variations of boundary surfaces and does not contribute to any forces. As discussed in the introduction this introduces a cutoff at a distance scale of $\Lambda^{-1}$, where $\Lambda$ is a frequency cutoff. In the regularization scheme used, this means that the minimum geodesic distance is chosen to be $\tilde{x}-x\equiv\Lambda^{-1}$. Inserting $\Lambda^{-1}$ as the lower limit for $r=0$ and summing over $r$, we obtain the contribution of odd paths $\epsilon_{odd}=\int\limits_{\Lambda^{-1}}^{\infty}\left\langle 0\left|T_{00}\right|0\right\rangle dz$. We note that the integrand leads to a quartic, $\Lambda^{4}$, divergence which results in a cubic, $\Lambda^{3}$, divergent Casimir energy. This contribution displays the expected cubic divergence for a scalar field obeying a Dirichlet boundary condition. Its finite contribution is canceled by the higher odd paths
\begin{equation}
\epsilon_{odd}= \int\limits_{0}^{a}\left\langle 0\left|T_{00}\right|0\right\rangle dz+\int\limits_{a}^{2a}\left\langle 0\left|T_{00}\right|0\right\rangle dz+\ldots+\int\limits_{ra}^{(r+1)a}\left\langle 0\left|T_{00}\right|0\right\rangle dz+\ldots.
\label{23}\end{equation}
Therefore the Casimir energy is given by the familiar result of eq. (\ref{16}).
%%%%%%%%%%%%%%%%%%%%%%%%%%%%%%%%%%%%%%%%%%%%%%%%%%%%%
\subsection{Permeable plates}                                
For an infinitely permeable boundary the use of closed paths renders the Casimir energy analysis transparent and rather trivial as we only need to replace the boundary conditions in (\ref{8}) with,
 \begin{equation}
 \Phi\left(x^{0},0,x^{2},x^{3}\right)=0\qquad
 \partial_{x^{1}}\Phi\left(x^{0},x^{1},x^{2},x^{3}\right)\left.\right|_{x^{1}=a}=0
 .\label{24}\end{equation}                            
Note that the reflection factor will be (+1) instead of (-1).  For the closed path of length $n_{e}a=2ra$, the reflection factor $\left(-1\right)^{r}\left(+1\right)^{r}=\left(-1\right)^{r}$ must be inserted in the summation (\ref{15}), so that the result is
  \begin{equation}
\left\langle0\right|T_{00}\left|0\right\rangle=\frac{m^2}{8\pi^2} \sum\limits_{r=1}^{\infty}\frac{\left(-1\right)^{r}K_2\left(2mra\right)}{r^{2}a^{2}},
\label{25}\end{equation}                                                                       
which is similar to the case of twisted scalar fields discussed in \cite{1}. In the massless limit, it leads to the energy per unit area
   \begin{equation}
\epsilon=\frac{1}{16\pi^{2}a^{3}}\sum\limits_{r=1}^{\infty}\frac{\left(-1\right)^{r}}{r^4}= -\frac{7}{8}\frac{1}{16\pi^{2}a^{3}}\zeta\left(4\right)=-\frac{7}{8}\frac{\pi^2}{1440a^3}.
\label{26}\end{equation}
%%%%%%%%%%%%%%%%%%%%%%%%%%%%%%%%%%%%%%%%%%%%%%%%%%%%%%%%%%%%%%%%%                                          
\section{Casimir energy in spacetimes with non-trivial topology} 
If we choose a periodic condition in $x^{2}$ direction with the periodicity equal to $b$, we get a cylindrical topology $\Sigma=R\times S\times R$ with the conditions
\begin{equation}\begin{array}{c}
\Phi\left(x^{0},0,x^{2},x^{3}\right)=\Phi\left(x^{0},a,x^{2},x^{3}\right)=0\\
\Phi\left(x^{0},x^{1},x^{2},x^{3}\right)=\Phi\left(x^{0},x^{1},x^{2}+sb,x^{3}\right)
,\label{27}\end{array}\end{equation}
and a non-zero vacuum energy is expected. The above introduced closed geodesic paths start at $x$ and reach $\tilde{x}_{ns}$ such that 
\begin{equation}
\left( \tilde{x}^{0}_{ns},\tilde{x}^{1}_{ns},\tilde{x}^{2}_{ns},\tilde{x}^{3}_{ns}\right)=\left\{
\begin{array}{c}
\left(x^{0},x^{1}+n_{e}a,x^{2}+sb,x^{3}\right)\\
\left( x^0,3x^1+\left(n_{o}-1\right)a,x^2+sb,x^3\right)\\
\left( x^0,-x^1+\left(n_{o}+1\right)a,x^2+sb,x^3\right)
\end{array}
\right.
\label{28}\end{equation}
and $n,s$ are integers. These paths can be classified into $C_{ns}$ classes, with $n$ points on the plates and $\left|{s}\right|$-circulations in the $x^{2}$ direction. Negative values of $s$ refer to those paths that circle in the opposite direction. We see that $G^{(1)}\left(x,\tilde{x}\right)$ is given by 
$G^{(1)}\left(x,\tilde{x}\right)=\sum\limits_{n,s}G^{(1)}\left(x,\tilde{x}_{ns}\right)$. The tangent vector to the geodesic paths at  $x$ and $\tilde{x}_{ns}$ are related through the equations $\tilde{\nabla}_{\mu}\sigma_{n_{e}}=-\nabla_{\mu}\sigma_{n_{e}}$ and (\ref{18}). The lengths of the paths in the class $C_{2r,s}$ is fixed by the pair $\left(r,s\right)$ while that of $C_{2r+1,s}$ varies from $\left[\left(2ra\right)^{2}+\left(sb\right)^{2}\right]^{\frac{1}{2}}$ to $\left[\left(2\left(r+1\right)a\right)^{2}+\left(sb\right)^{2}\right]^{\frac{1}{2}}$. The zero path length belongs to the high energy fields in the class $C_{0,0}$. Like the previous example, in the space integration we have 
\begin {equation}
\epsilon_{2r+1,s}=\int\limits_{\left[\left(2ra\right)^{2}+\left(sb\right)^{2}\right]^{\frac{1}{2}}}^{\left[\left(2\left(r+1\right)a\right)^{2}+\left(sb\right)^{2}\right]^{\frac{1}{2}}}\left\langle 0\left|T_{00}\right|0\right\rangle dz,
\label{29}
\end{equation}
and for the class $C_{1,0}$, the distance scale $\Lambda^{-1}$ enters at the lower limit and the finite contributions are canceled by the higher $C_{2r+1,s}$ classes. We get the finite results from the paths in $C_{2r,s}$ with $\left(r,s\right)\neq\left(0,0\right)$. In this topology, the Casimir energy per unit area is
\begin{equation}
\epsilon=\frac{m^{4}a}{4\pi^{2}} \sum\limits_{r,s=-\infty,\left(r,s\right)\neq\left(0,0\right)}^{\infty}
\frac{K_2\left(z_{rs}\right)}{z_{rs}^{2}},
\label{30}\end{equation} 
where $z_{rs}=m\left[\left(2ra\right)^{2}+\left(sb\right)^{2}\right]^{\frac{1}{2}}$. 
Our result in (\ref{30}) would be in agreement with that obtained in \cite{7} for $\Sigma=S^{2}\times R$, if the periodic length in $x^1$ direction is chosen to be $2a$. Later we will use the analogy between the above Casimir energy calaculation in a nontrivial toplogy and that at finite temperature, to derive the free energy of a cavity at finite temperature. Using similar arguments for a spacetime with periodicities $b$ and $c$ in the $x^{2}$ and $x^{3}$ directions and the boundary conditions
 \begin{eqnarray}\begin{array}{c}
\Phi\left(x^{0},0,x^{2},x^{3}\right)=\Phi\left(x^{0},a,x^{2},x^{3}\right)=0\\
\Phi\left(x^{0},x^{1},x^{2},x^{3}\right)=\Phi\left(x^{0},x^{1},x^{2}+sb,x^{3}+lc\right)
,\label{31}\end{array}\end{eqnarray} we obtain
\begin{equation}
\epsilon=\frac{m^{4}}{4\pi^{2}} \sum\limits_{r,s,l=-\infty,\left(r,s,l\right)\neq\left(0,0,0\right)}^{\infty}\frac{K_2\left(z_{rsl}\right)}{z_{rsl}^2}
\label{32}\end{equation}                                                                                    where $z_{rsl}=m\left[\left(2ra\right)^{2}+\left(sb\right)^{2}+\left(lc\right)^{2}\right]^{\frac{1}{2}}$. It can be seen that $z_{rs}$ and $z_{rsl}$ are nothing but the mass times the length of classical periodic path. A typical non-planar path starting in a given direction includes $\left|2r\right|$-reflections from the surfaces $x^{1}=0,a$, $\left|s\right|$-circulations in the $x^{2}$ direction and $\left|l\right|$-circulations in the $x^{3}$ direction. Again negative values of $s$ and $l$ refer to those paths that circle in the opposite direction. A factor of two in $2ra$ term is due to the Dirichlet boundary condition in the $x^{1}$ direction. If the periodic condition in $x^{2}$ direction is replaced with a Dirichlet boundary condition, we get the similar Casimir energy expression as in eq. (\ref{32}) but now with   
$z_{rsl} = m\left[\left(2ra\right)^{2}+\left(2sb\right)^{2}+\left(lc\right)^{2}\right]^{\frac{1}{2}}$.
The Neuman boundary condition that the normal derivative of the field, rather than the field itself vanishes at the surface leads to the insertion of either $(-1)^{n}$ or $(-1)^{s}$ for each permeable wall in $x^{1}$ or $x^{2}$ direction respectively.\\
Since the fields at $x$ and $\tilde{x}_n$ ($n$ a family index) are related through $\Phi\left(x\right)=\left(-1\right)^{n}\Phi\left(\tilde{x}_{n}\right)$,
applying the optical approach to twisted fields, different paths appear in eq. (\ref{10}) with non-equal weights. So a twist in any specific direction inserts a factor of $\left(-1\right)^{n}$, where $n$ is the number of circulations in that direction. 
%%%%%%%%%%%%%%%%%%%%%%%%%%%%%%%%%%%%%%%%%%%%%%%
\subsection{Non-orientable manifolds}
The manifolds we have examined so far are orientable. Here we examine the effect of spacetime manifold  orientation on the closed geodesic paths and consequently on the vacuum polarization. Since frames could only be defined locally and transport around closed spatial directions may change the handedness of local frames, it is expected that the non-orientability of a manifold could affect the Casimir energy calculations. A four dimensional Mobius strip $R^1\times \Sigma = R^{1}\times M^2\times R^1$ is a non-orientable spacetime manifold which can be constructed by the identification of the points  $\left(x^0,x^1,x^2,x^3\right)$ and $\tilde{x}_n=\left(x^0,x^1+na,\left(-1\right)^{n}x^2,x^3\right)$ in Minkowski spacetime. So for a scalar field in this manifold we have
\begin{equation}
\Phi\left(x^0,x^1,x^2,x^3\right)= \Phi\left(x^0,x^1+na,\left(-1\right)^{n}x^2,x^3\right). 
\label{33}\end{equation}
Different paths in class $C_{n}$, circulate $\left|n\right|$-times around $x^{1}$ direction and reach $\tilde{x}_n$ with the half squared geodesic lengths equal to,
\begin{equation}\left\{ \begin{array}{l}   \sigma_{2r}\\\sigma_{2r+1}\end{array}\right.=\frac{1}{2}\left[\left(x^{0}-\tilde{x}^{0}\right)^2-\left\{\begin{array}{c}\left(x^{1}-\tilde{x}^{1}-2ra\right)^2+\left(x^{2}-\tilde{x}^{2}\right)^2\\\left(x^{1}-\tilde{x}^{1}-\left(2r+1\right)a\right)^2+\left(x^{2}+\tilde{x}^{2}\right)^2\end{array}\right\}-\left(x^{3}-\tilde{x}^{3}\right)^2 \right].            
  \label{34}\end{equation} 
The lengths of the paths in classes $C_{2r}$ and $C_{2r+1}$ are given by $\left|2ra\right|$ and $\left[\left(2r+1\right)^{2}a^{2}+\left(2x^{2}\right)^{2}\right]^\frac{1}{2}$ respectively. The zero-length paths fall into $C_{0}$ class and so the finite contribution to the Casimir energy is due to the other classes. Now using the facts that,
 \begin{equation}     
\tilde{\nabla}_{\mu}\sigma_{2r}=-\nabla_{\mu}\sigma_{2r}\qquad
\tilde{\nabla}_{\mu}\sigma_{2r+1}=-\nabla_{\mu}\sigma_{2r+1}+2\delta_{\mu,2}\nabla_{1}\sigma_{2r+1},
\label{35}\end{equation}
we have
 \begin{eqnarray}
\left\langle0\left|T_{00}\right|0\right\rangle&=&
   -\frac{1}{2}\lim_{\tilde{x}\rightarrow x}\sum_{r=1}^{\infty }\left[2\nabla_{0}\nabla_{0}G^{(1)}\left(\sigma_{2r}\right)+\left[\nabla_{0}\nabla_{0}-\left(4\varsigma -1\right)\nabla_{2}\nabla_{2}\right]G_{ren}^{(1)}\left(\sigma_{2r+1}\right)\right]\nonumber\\  &=&\frac{m^4}{2\pi^2}\sum\limits_{r=1}^{\infty}\frac{K_2\left(A_{2r}\right)}{\left(A_{2r}\right)^2}+\frac{m^4}{4\pi^2}\sum\limits_{r=-\infty}^{\infty}\nonumber\\
   &&\left[2\left(1-2\varsigma\right)\frac{K_2\left(A_{2r+1}\right)}{\left(A_{2r+1}\right)^2}
 +\left(4\varsigma-1\right)\left(2x^{2}m\right)^{2}\frac{K_3\left(A_{2r+1}\right)}{\left(A_{2r+1}\right)^3}\right],
\label{36}\end{eqnarray}
where $A_{2r}=2mra$ and $A_{2r+1}=m\left[\left(2r+1\right)^{2}a^{2}+\left(2x^{2}\right)^{2}\right]^\frac{1}{2}$. This expression has been obtained in \cite{7} using the method of images.
%%%%%%%%%%%%%%%%%%%%%%%%%%%%%%%%%%%%%%%%%%%%%%%%%%
\subsection{Mass and free energy expansion}
The above classical closed paths formalism could also be employed to calculate the {\it free energy}
of the cavity discussed in the previous section at finite temperature \cite{13}. To do so
one could use the well known fact in the field theory at finite tempertaure that a system in three dimensional space in equilibrium at temperature $T$ could be treated as a system in a four-dimensional Euclidean space whose fourth dimension is compactified to a circle of circumference $l_{T}=\frac{1}{k_{B}T}$.
This amounts to replacing in (\ref{30}) the contribution of each closed path of class $C_{n}$ at $T=0$ by the sum of contributions of closed paths, now with added $\left|s\right|$-circulations in the extra fourth dimension. This extends the path lengths from $l_{n}$ to $l_{n}^{T}=\left[l_{n}^2+\left(sl_{T}\right)^2\right]^{\frac{1}{2}}$. So, using (\ref{30}) the free energy per unit area $\it F$ is given by
 \begin{equation}
{\it F}=\frac{m^{4}a}{2\pi^2}\sum_{n=1}^{\infty}\sum_{s=-\infty}^{\infty}\frac{K_{2}\left(l_{n}^T\right)}
{\left(l_{n}^T\right)^2}.
\label{37}\end{equation}
Note the difference between eqs. (\ref{30}) and (\ref{37}). When the boundary conditions (\ref{27}) are applied, the eigenmodes will be labeled with two quantum numbers and consequently closed paths will be classified in classes labeled with two indices and, as before, zero length paths should be ignored. 
On the other hand for $T > 0$, the eigenfrequencies of the cavity change from $\omega_n$ to $\omega_{n}^{\prime}$ while the quantum numbers of a mode are conserved \cite{13}. So only the paths in classes $C_{n\neq 0}$ will contribute while their lengths are given by $l_{n}^{T}$ instead of $l_{n}$. In other words, there are no contributions in (\ref{37}) due to classes $C_{ n= 0 , s\neq 0}$ and the finite temperature does not induce any new ultraviolet divergences.
Using (\ref{37}) the free energy density can be written in the following form 
\begin{equation}
{\cal{F}}=\frac{m^{4}}{2\pi^2}\sum^{\infty}_{n=1}\sum_{s=-\infty}^{\infty}\frac{K_2\left(l_{n}^{T}\right)}{\left(l_{n}^{T}\right)^2},
\label{38}\end{equation}
where its high and low temperature expansions can be obtained by introducing the variables $\tau=2ak_{B}T$ and $\alpha=m\sqrt{2a/k_{B}T}$.
 $l_{n}^{T}$  may be rewritten as $l_{n}^T=\alpha\left(n^{2}\tau+\frac{s^2}{\tau}\right)^\frac{1}{2}$, and the summation $\sum\limits_{s=-\infty}^{\infty}\frac{K_2\left(l_{n}^T\right)}{\left(l_{n}^T\right)^2}$  is found by using the identity

$$\sum\limits_{n=-\infty}^{\infty}f\left(n\right)= - {\rm the\;sum\;of\;residues\;of\;} \pi \cot\left(\pi z\right)f\left(z\right) {\rm at\;all\;poles\;of\;} f\left(z\right). $$
%\noindent 
In our case there are two poles at $s=\mp in\tau$ and the residue at $s=in\tau$ is given by
\begin{equation}
\frac{\pi^2}{2\alpha^{4} n^2}\left[1-\coth^{2}\left(\pi \tau n\right)\right]-\frac{\pi \coth\left(\pi \tau n\right)}{2\alpha^{4}n^{3}\tau}+\frac{\pi \coth\left(\pi \tau n\right)}{4\alpha^2n}
.\label{39}\end{equation}                                                
As this is symmetric under $n\longleftrightarrow -n$, the residue at $s=-in\tau$ is the same and we obtain 
\begin{equation}
\sum_{s=-\infty}^{\infty}\frac{K_{2}\left(l_{n}^{T}\right)}{\left(l_{n}^{T}\right)^2}=-\frac{\pi^2}{\alpha^{4} n^2}\left[1-\coth^{2}\left(\pi \tau n\right)\right]+\frac{\pi \coth\left(\pi \tau n\right)}{\alpha^{4}n^{3}\tau}-\frac{\pi \coth\left(\pi \tau n\right)}{2\alpha^2 n}
.\label{40}\end{equation}                            
It could be seen that after multiplicatiom by ${m^4}/{2\pi^2}$ and summation over $n$, only the last term depends on 
the mass field and its contribution reduces the energy density. It is also seen that the mass term is completely 
separable and divergent in large $\tau$ limit. It also behaves like $\sum\limits_{n=1}^{\infty}n^{-1}$ which introduces 
a new divergence. For massless fields and large $\tau$ limit it corresponds to 
\begin{equation}
\sum_{s=-\infty}^{\infty}\frac{K_{2}\left(l_{n}^{T}\right)}{\left(l_{n}^{T}\right)^2}=
\frac{4\pi^2}{\alpha^{4}n^2}\sum\limits_{k=1}^{\infty}k e^{-2\pi n\tau k}+\frac{\pi}{\alpha^{4}n^{3}\tau}
\left[1+2\sum\limits_{k=1}^{\infty}e^{-2\pi \tau nk}\right]
,\label{41}\end{equation}                                       
and the free energy density is given by
\begin{eqnarray}
{\cal{F}}&=&\frac{1}{2\pi ^{2}}\left[\frac{\pi}{\tilde{\alpha} ^{4}\tau}\zeta\left(3\right)+\frac{2\pi}
{\tilde{\alpha} ^{4}\tau}\sum\limits_{k=1}^{\infty}\left(1+2\pi \tau k\right) e^{-2\pi \tau k} \sum\limits_{n/k = integer>0} n^{-3}\right]\nonumber\\
&=&\frac{\zeta\left(3\right)k_{B}T}{16\pi a^{3}}+\frac{k_{B}T}{8\pi a^{3} }\left(1+4\pi ak_{B}T\right) e^{-4\pi ak_{B}T}+O \left(e^{-8\pi ak_{B}T}\right).
\label{42}\end{eqnarray}                                                                        
where $\zeta \left(3\right)=\sum\limits_{n=1}^{\infty}n^{-3}$ and $\tilde{\alpha}=\alpha/m=\sqrt{2a/{k_{B}T}}$.
Free energy at high temperature (fixed $a$) converges like $\sum\limits_{n=1}\frac{1}{n^3}$. Although its convergence is slower than that in the case at $T=0$, the contribution of paths with small number of reflections is still a good apprroximation. To study the small $\tau$ (low temperature) behaviour of the free energy density we first note a reflection symmetry $\tau\rightarrow 1/\tau$ in ${\cal{F}}-{m^4\over 2\pi^2}\sum\limits_{n=1}^{\infty}\frac{k_{2}\left(\alpha n\sqrt{\tau}\right)}{\alpha^{2}n^{2}\tau}$ which  relates the high and low temperature regimes. So with $\tau$ replaced by $1/\tau$ in this expression, it is suitably adapted for obtaining the small $\tau$ limit. Doing so we get another expression for eq. (\ref{38}) in the following form 
\begin{eqnarray}
{m^4\over 2\pi^2}\sum\limits_{n=1}^{\infty}\left[-\frac{\pi^2}{\alpha^{4} n^2}\left[1-\coth^{2}\left(\pi n/\tau\right)\right]+\frac{\pi\tau \coth\left(\pi n/\tau\right)}{\alpha^{4} n^3}\right.\nonumber\\-\left.\frac{\pi\coth\left(\pi n/\tau\right)}{2\alpha^{2}n}+\frac{K_2\left(\alpha n\sqrt {\tau}\right)}{\alpha^{2} n^{2}\tau}-\frac{K_2\left(\alpha n/\sqrt {\tau}\right)}{\alpha^{2} n^{2}/\tau}\right]
.\label{43}\end{eqnarray}
The last three terms show the mass dependence. Although from the previous studies \cite{7}-\cite{8} it is known that 
the contribution of massive fields  can be neglected relative to their massless counterpart, here infinities show up 
in low as well as in high temperature regimes whenever the fields are massive. The zero-point energy of a confined 
massive scalar field in two dimensions is studied in \cite{15} (see also \cite{16}) where the mass dependent logarithmic 
divergence is absorbed into a redefinition of the so called {\it bag constant}. But in our case it is a problematic 
divergence as it can not be absorbed into a constant. In the massless case and for $\tau \ll  1$ we have
\begin{eqnarray}
&&{\cal{F}}=\frac{\pi^2}{1440a^{4}}-\frac{\pi^2\left(k_{B}T\right)^4}{90}+\frac{\zeta\left(3\right)\left(k_{B}T\right)^3}{4\pi a}+\frac{\left(k_{B}T\right)^3}{\pi a}\left(1+\frac{\pi }{ak_{B}T}\right)e^{-\pi /ak_{B}T}\nonumber\\&+&O \left(e^{-2\pi /ak_{B}T}\right)
%\nonumber\\&=&
={\cal{F}}_{0}\left[1-\left(\frac{T}{T_{eff}}\right)^{4}+\frac{45\zeta\left(3\right)}{\pi^3}\left(\frac{T}{T_{eff}}\right)^{3}\right. \nonumber\\ 
&&\left.+\frac{720}{\pi^3}\left(\frac{T}{T_{eff}}\right)^{3}\left(1+2\pi\frac{T_{eff}}{T}\right)e^{-2\pi\frac{T_{eff}}{T}}+O\left(e^{-4\pi\frac{T_{eff}}{T}}\right)\right]
.\label{44}\end{eqnarray}
Here ${\cal{F}}_{0}$ is the zero-temperature Casimir energy density and  $T_{eff}=\frac{1}{2ak_{B}}$ 
is the effective temperature. The results (\ref{42}) and (\ref{44}) can be compared with the high and low temperature 
expansions obtained in \cite{13} and \cite{17}. Presence of the term proportional to $T^4$ in low $T$-expansion 
is an expected result when we note that it has
the same form as the free energy density of black body radiation of the same temperature in the same volume. The 
appeareance of this term originates from the fact that, as in the case of zero temperature, in renormalization of the 
free energy  of the cavity we need to subtract from it the free energy of the unbounded empty space at the given 
temperature in the same volume (For a detailed discussion on this refer to \cite{2} and references therein).

%%%%%%%%%%%%%%%%%%%%%%%%%%%%%%%%%%%%%%%%%%%%%%%%%%%%%%%%%%%%%%
\section{Optical approach and spinor fields}
In \cite{8} we have discussed different spin structures, twisted and untwisted, on  spacetime manifolds with non-trivial topologies by choosing different local frames. Analogous to eq. (\ref{10}), for spinor fields we have  
\begin{equation}
\left\langle 0\left|T^{\mu\nu}\right|0\right\rangle={\frac{1}{16}}\lim_{\tilde{x}\rightarrow
x} \mathrm{Tr}\left[(\gamma^{\mu}\nabla^\nu + \gamma^{\nu}\nabla^\mu) -
(\gamma^{\mu}\tilde{\nabla}^\nu + \gamma^{\nu}\tilde{\nabla}^\mu)\right]
\gamma^\sigma \nabla_\sigma \sum_{n}G_{ren}^{(1)}(x,\tilde{x}_{n})
.\label{45}\end{equation}
For untwisted spinor fields and orientable manifolds, eq. (\ref{45}) reduces to $\left\langle 0\left|T_{\mu\nu}\right|0\right\rangle=\lim_{\tilde{x}\rightarrow x} \nabla^{\mu}\nabla^{\nu}G^{(1)}\left(x,\tilde{x}\right)$.  Making use of the optical approach, different closed paths appear with the same weight, except the zero length ones which should be discarded. Like the scalars, for twisted spinors the factor of $\left(-1\right)^{n}$ enters in eq. (\ref{45}) to show the twist of the frame in the direction with $n$-circulation.
In a non-orientable manifold such as the Mobius strip, we will see that limited number of paths contribute due to the fact that the local frames at $x$ and $x_{n}$ (which are related through a Lorentz transformation) are not the same. The spinor $\Psi\left(x_{n}\right)$ is related to $\Psi\left(x\right)$ through $\Psi\left(x_{n}\right)=S\Psi\left(x\right)$, where $S$ is a suitable representation of $SL\left(2,C\right)$. This translates into the Casimir energy expansion in the form of different weights for different closed paths. In Mobius strip, the local frames at $x$ and $x_{n}$ are related through parity and $S=\left(\gamma^{0}\gamma^{3}\gamma^{1}\right)^{n}$ such that $S^{2}=I$ and therefore,
\begin{equation}
G_{ren}^{(1)}(\sigma) = \sum_{n=-\infty, n \neq 0}^\infty \left[
G_0^{(1)}(\sigma_{2n}) +
G_0^{(1)}(\sigma_{2n+1})(\gamma^0\gamma^3\gamma^1)\right]
.\label{46}\end{equation}
Since the trace of an odd numbers of $\gamma$-matrices vanishes, taking the trace, reveals that the second sum in eq. (\ref{46}) makes no contribution and we immediately gain
\begin{equation}
\left\langle0\left|T_{00}\right|0\right\rangle=-\frac{m^4}{\pi^2}\sum\limits_{r=1}^{\infty}\frac{K_2\left(2mra\right)}{\left(2mra\right)^2}
,\label{47}\end{equation}
which, as expected, is minus the twice of the first term in eq. (\ref{36}).  
%%%%%%%%%%%%%%%%%%%%%%%%%%%%%%%%%%%%%%%%%5
\section{Conclusions}
Following Casimir's original prediction for the force between grounded conducting plates, many other 
calculations have been carried out for different variations in the conductor geometry or background 
\cite{18,19} and many exact and approximate techniques have been proposed to calculate this energy. 
Here as the main objective, employing the recently proposed optical approach, we calculated the Casimir 
energy for massive fields in a  cylindrical topology as well as
on a Mobius strip as a non-orientable manifold. The 
simplification provided by this approach allows the extension of the domain of applicability to 
the calculation of the regularized vacuum energy of a quantum field subjected to various static 
external conditions. Applying this approach to systems at finite temperature, the finite temperature 
effect on the free energy of a cavity is compared with that of a periodic boundary condition at zero 
temperature. The same approach has also been applied to calculate the Casimir energy for spinors and 
twisted fields, where it is shown that different paths contribute with non-equal 
weight factors. Using the same approach we have recovered the well known fact that the dependence of 
free energy on mass at high and low temperatures may be regarded as problematic since the key feature 
here is the presence of a divergent term which is absent in the massless case. As a consequence the 
contribution of massive fields in these regimes not only does not tend to damp relative to their 
massless counterparts (as in the $T=0$ case) but also diverges. 
%%%%%%%%%%%%%%%%%%%%%%%%%%%%%%%%%%%%%%%%%%%%
\section *{Acknowledgments} 
The authors would like to thank University of Tehran for supporting this project under the grants 
provided by the research council.
%%%%%%%%%%%%%%%%%%%%%%%%%%%%%%%%%%%%%%%%%

\end{document}